\documentstyle[11pt,epsf]{article}
\begin{document}

\begin{titlepage}
\samepage{
\setcounter{page}{1}

\rightline{UFIFT--HEP--97--15}
\vfill
\begin{center}
 {\Large \bf Spin-Dependent Drell-Yan and Double Prompt Photon Production 
to NLO QCD $\dagger$}

\vfill
\vfill
 {\large Sanghyeon Chang$^{*}$\footnote{
        E-mail address: schang@phys.ufl.edu},
        Claudio Corian\`{o},$^{*}$\footnote{E-mail address: coriano@phys.ufl.edu}   
        R. D. Field$^{*}$\footnote{
        E-mail address: rfield@phys.ufl.edu}
        \\$\,$ and  L. E. Gordon$^{\dagger\dagger}$ \footnote 
{E-mail address: gordon@hep.anl.gov}}
\\
\vspace{.12in}
 {\it $^{*}$   Institute for Fundamental Theory, Department of Physics, \\
        University of Florida, Gainesville, FL 32611, 
        USA\\}

\vspace{.075in}

{\it  $^{\dagger\dagger}$ Argonne National Laboratory\\
9700 South Cass Av, IL USA\\}
\end{center}
\vfill
\begin{abstract}

We present the complete $O(\alpha_s^2)$ radiative corrections to the 
(non singlet) polarized Drell Yan cross section for the production of a lepton pair with a nonzero $q_T$. The helicity of the incoming states is arbitrary. 
In the case of double photon,
results for the longitudinal asymmetries and on the $p_T$ behavior of the 
cross section are also given  
(to $O(\alpha_{em}^2\alpha_s )$) in the central rapidity region of the tagged photon.

\end{abstract}
\smallskip}
\vspace{0.5cm}
$\dagger${Presented by C. Corian\`{o} 
at the Fifth International meeting DIS 97,\\ Chicago 14-18 April 1997 }
\end{titlepage}

\setcounter{footnote}{0}

\def\beq{\begin{equation}}
\def\eeq{\end{equation}}
\def\beqn{\begin{eqnarray}}
\def\eeqn{\end{eqnarray}}

\def\ie{{\it i.e.}}
\def\eg{{\it e.g.}}
\def\half{{\textstyle{1\over 2}}}
\def\third{{\textstyle {1\over3}}}
\def\quarter{{\textstyle {1\over4}}}
\def\m{{\tt -}}

\def\p{{\tt +}}

\def\slash#1{#1\hskip-6pt/\hskip6pt}
\def\slk{\slash{k}}
\def\GeV{\,{\rm GeV}}
\def\TeV{\,{\rm TeV}}
\def\y{\,{\rm y}}

\def\l{\langle}
\def\r{\rangle}

\setcounter{footnote}{0}
\newcommand{\beqa}{\begin{eqnarray}}
\newcommand{\eeqa}{\end{eqnarray}}
\newcommand{\eps}{\epsilon}


\setcounter{footnote}{0}

\section{Drell Yan Lepton Pair Production at Nonzero $q_T$} 

Spin physics is a new avenue for QCD. While most of studies 
at hadron colliders
have been focused in the past on the analysis of unpolarized collisions, more recently a large research activity in QCD has shifted toward a study of the spin structure of the nucleon in DIS at HERA, while new experiments have 
been planned for investigating polarized $p-p$ collisions at RHIC and at 
HERA-N. 

RHIC is supposed to run in between 50-500 GeV center of mass energy, and, 
as a p-p collider, will have a very high luminosity ($500$ pb$^{-1}$). Therefore,we have the hope to be able to gather from this experiments enough 
evidence in order to better describe the flavor and the $\Delta g$ 
contribution in the nucleon over a wide range of $x$ ($\equiv x_{BJ}$) and $Q^2$. 
Given the fact that the asymmetries are usually small, unless one looks 
at specific intervals of $x$ and $ Q^2$, 
it is important to focus the investigations over those processes which promise to generate large asymmetries and have 
a particularly clean final state. Among these processes, single and 
double prompt photon production together with the Drell-Yan process 
are surely among the most interesting to be studied. 

All the models which have been proposed for the polarized structure 
functions \cite{GS,GRSV}, the $\Delta g$ content appears to be not well 
determined 
(for instance at $x< 0.1$ and $Q^2\sim 100$ GeV$^2$, see ref.~\cite{CG} 
for details).

The Drell-Yan process will be crucial in order to study the distribution of transverse spin in the nucleon and unveil the transversity distribution $h_1$, since the gluons decouple at leading order. 

We have studied the non singlet contributions to the longitudinally 
polarized Drell-Yan cross section to $O(\alpha_s^2)$ \cite{CCFG}. 
The calculation is done for a nonzero $q_T$ of the lepton pair and starts 
(at Born level) at order $\alpha_s$. In the unpolarized case, 
this cross section was studied by Ellis Martinelli and Petronzio \cite{EMP}.

The integration of our results in $t-u$ 
in dimensional regularization gives the bulk of the NNLO corrections to the invariant mass distributions ${d\sigma/d Q^2}$ of the lepton pair. 

At small $p_T$ the cross section needs resummation. To leading order 
this has been done in ref.~\cite{weber} following the work of refs. \cite{alt,ster}. 

It is important to extend completely these results to the polarized 
case both in the case of longitudinal and transverse asymmetries.

The next to leading order ($\alpha_s$) corrections to Drell-Yan total cross section \
are quite simple, since the 
process starts at $O(1)$ through the annihilation channel, 
and well known \cite{ratcliffe}. They have been reconsidered by various 
authors (see \cite{CCFG} and refs. therein).  
They involve the usual vertex and 
self-energy corrections to the lowest order process ($q\bar{q}\gamma$) 
and can be calculated quite straightforwardly. 

The parton level analysis on the nonzero $q_T$ distributions is very 
involved and, unfortunately, requires a recalculation of all the radiative 
corrections -including the virtual corrections- considered before in ref. 
\cite{EMP}. Our calculation is the first complete check of \cite{EMP}, since 
the helicity is an additional parameter that can be sent to zero at any stage of the calculation. 

We decided to use the $\overline{MS}$ scheme since the 
evolution equations are also treated in the same scheme. We show that in this scheme helicity is not conserved and one needs a finite renormalization to 
reenter into a physical scheme. 

The non singlet sector of Drell-Yan is defined as a difference between 
scattering and annihilation diagrams of quarks and antiquarks. 

 At the parton level we define 
\beqa
&& \sigma_{NS}\equiv\sum_{f,f'}(\sigma_{q_f \bar{q}_{f'}} - \sigma_{q_f
q_{f'}}), 
\eeqa
where we have included a sum over all the flavors $f, f'$. 
A careful diagrammatic analysis shows that the cross section so defined is 
diagonal in flavor.

We refer to our works for more details. Here we just quote the result. 
In the $\overline{MS}$, to order $\alpha_s^2$ the non-singlet parton level spin asymmetry becomes
\begin{eqnarray}
s{d{\hat\sigma}^{LL}_{NS}\over dtdu}(s,t,u)=
-s{d{\hat\sigma}^\Sigma_{NS}\over dtdu}+
s{d{\hat\sigma}^{hat}_0\over dtdu}-
2s{d{\hat\sigma}^\Sigma_{4}\over dtdu},
\label{rdf_nspin1}
\end{eqnarray}
where $d{\hat\sigma}^\Sigma_{NS}$ is the non-singlet unpolarized spin averaged cross section of ref. \cite{EMP}, while $d\hat\sigma^{hat}_0$ is a regularization scheme dependent helicity non-conserving piece. 
Similarly, ${d{\hat\sigma}^\Sigma_{4}}$ is the unpolarized contributions 
coming from the $qq$ scattering sector.

In this scheme $P_{qq}(z)=\Delta P_{qq}(z)$ in $D$-dimensions and the helicity 
violating terms are $explicitely$ needed in the convolutions 
of the hard scattering with the NLO evolved structure functions. 
The evolution DGLAP kernel has obviously to be evaluated in this same scheme.

\section{Double Photon Production in Polarized Collisions to NLO}
The analysis of Double Photon production can be found in ref.~\cite{CG}.
Here we quote briefly some of the results.

\begin{figure}
\epsfxsize=60mm
\centerline{\epsfbox{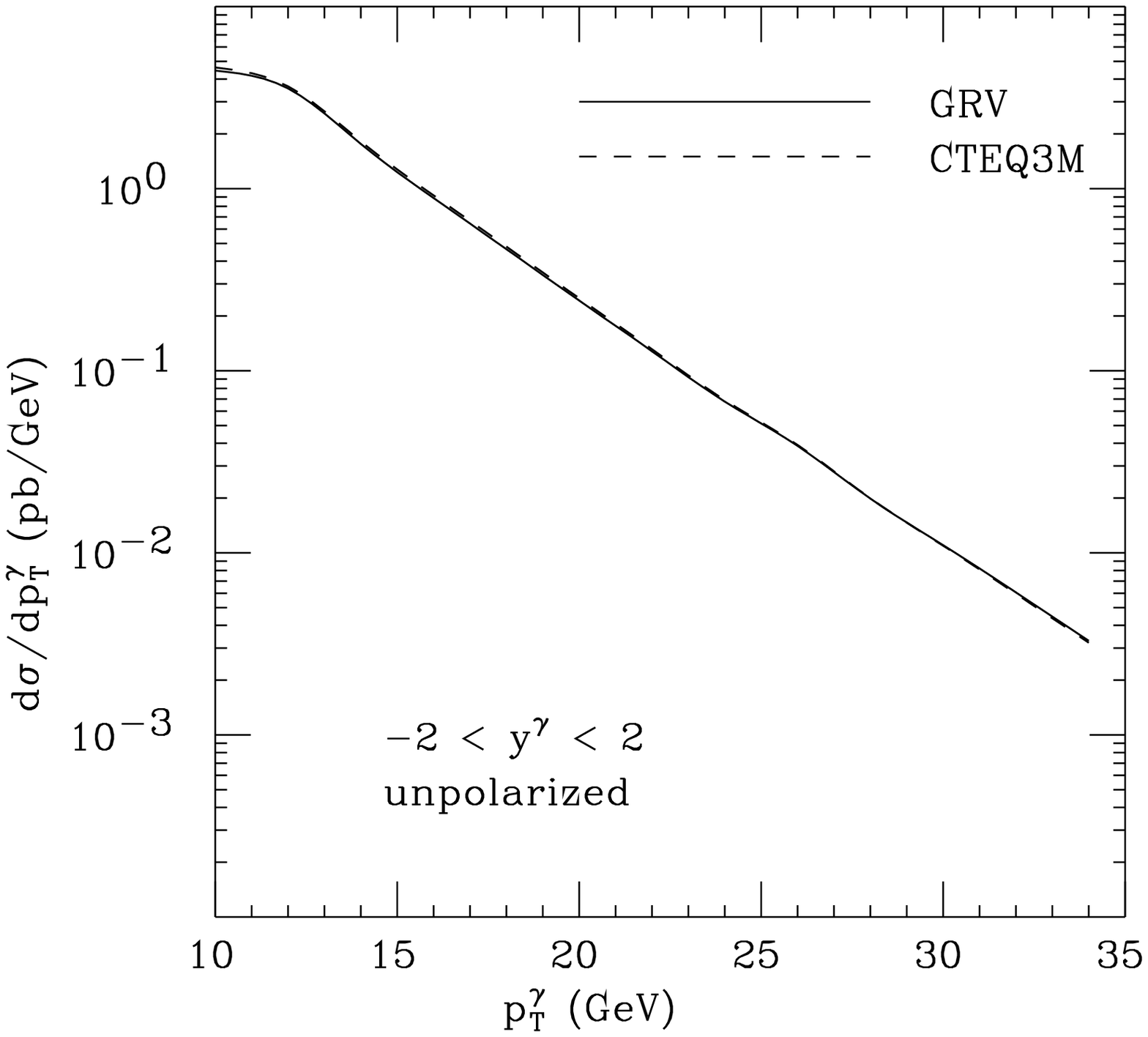}
\epsfxsize=60mm
\epsfbox{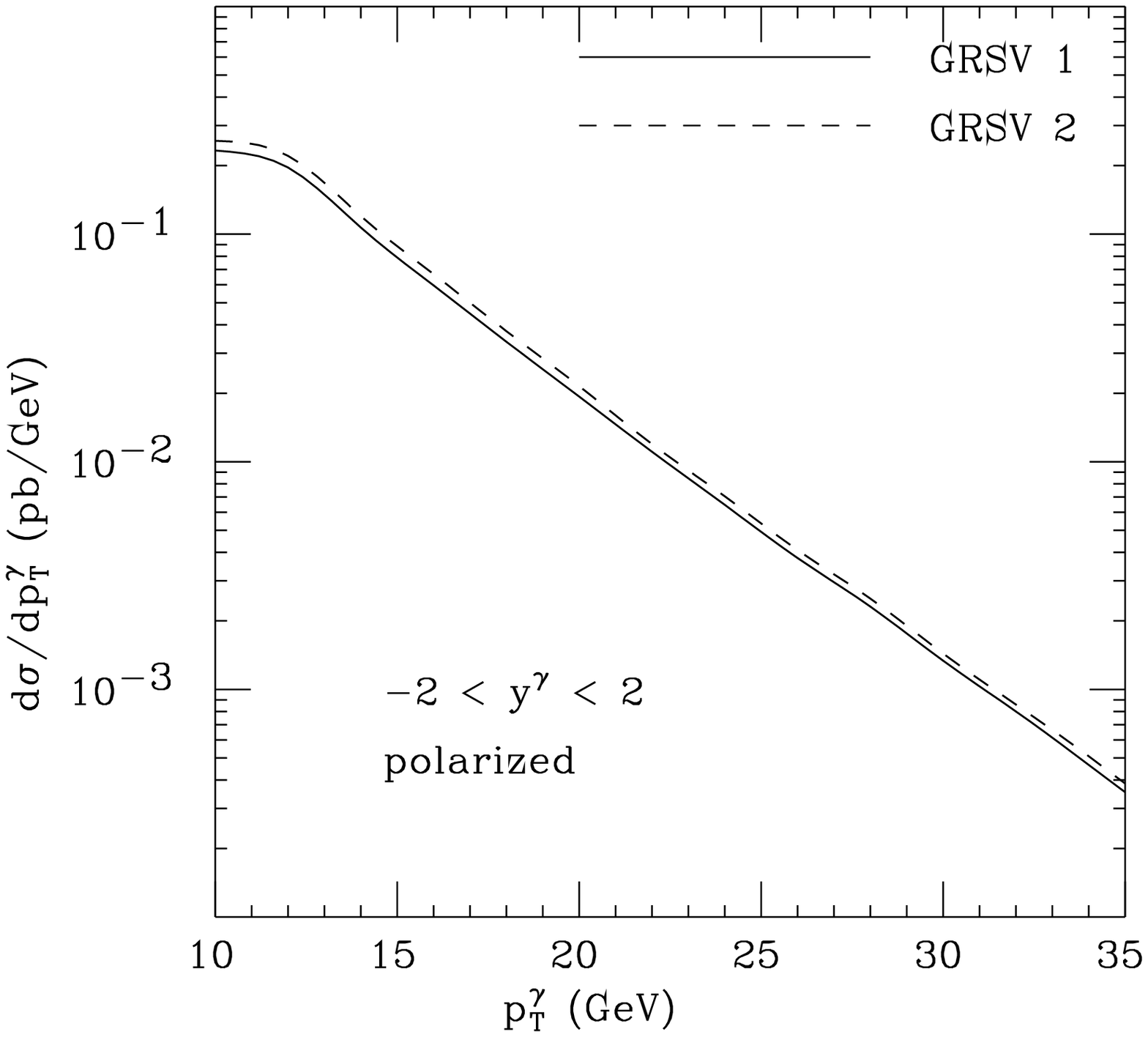}}
\centerline{a)\hspace{6cm} b)}
\caption{}
\end{figure}
\begin{figure}
\epsfxsize=60mm
\centerline{\epsfbox{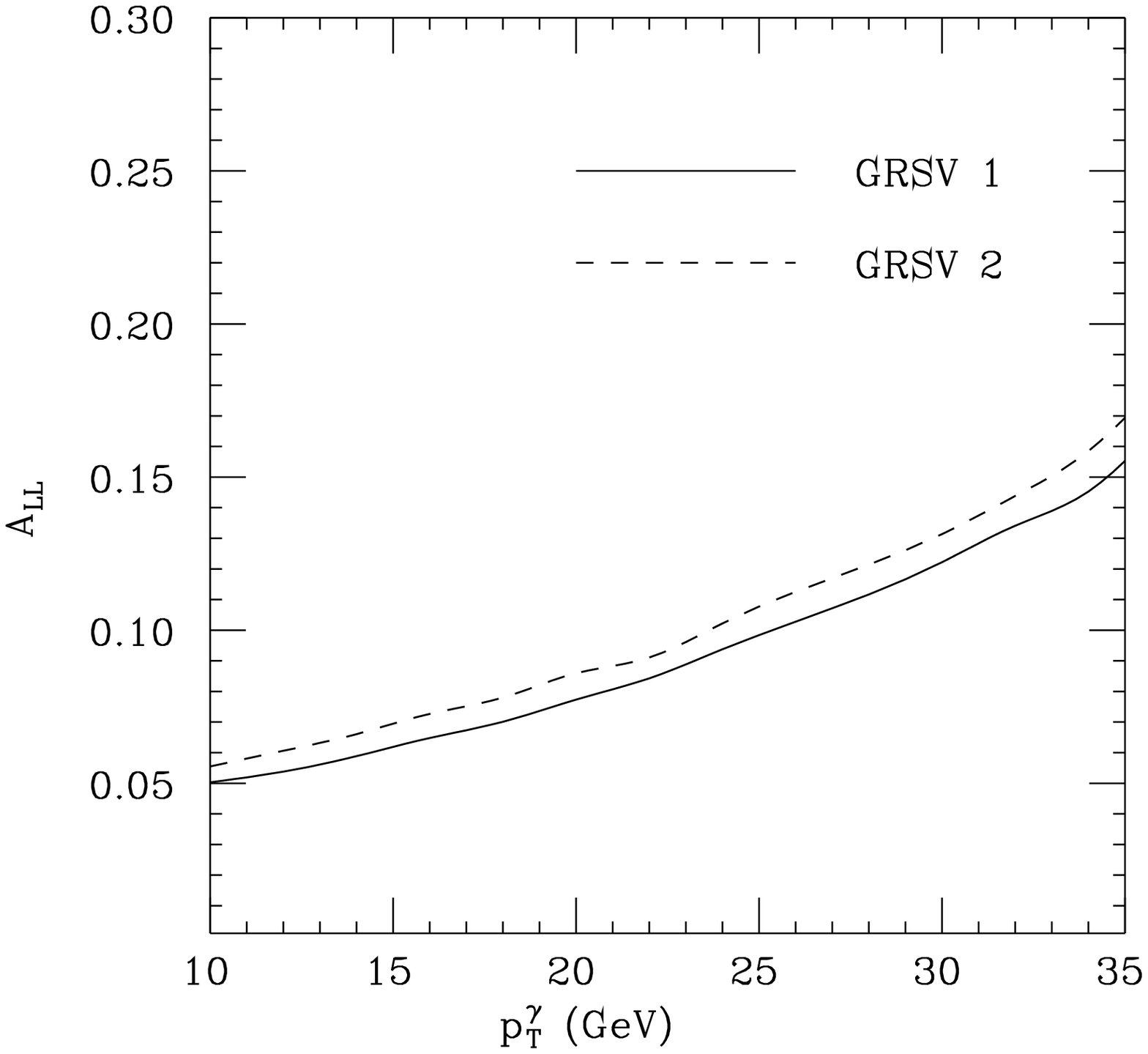}
\epsfxsize=60mm
\epsfbox{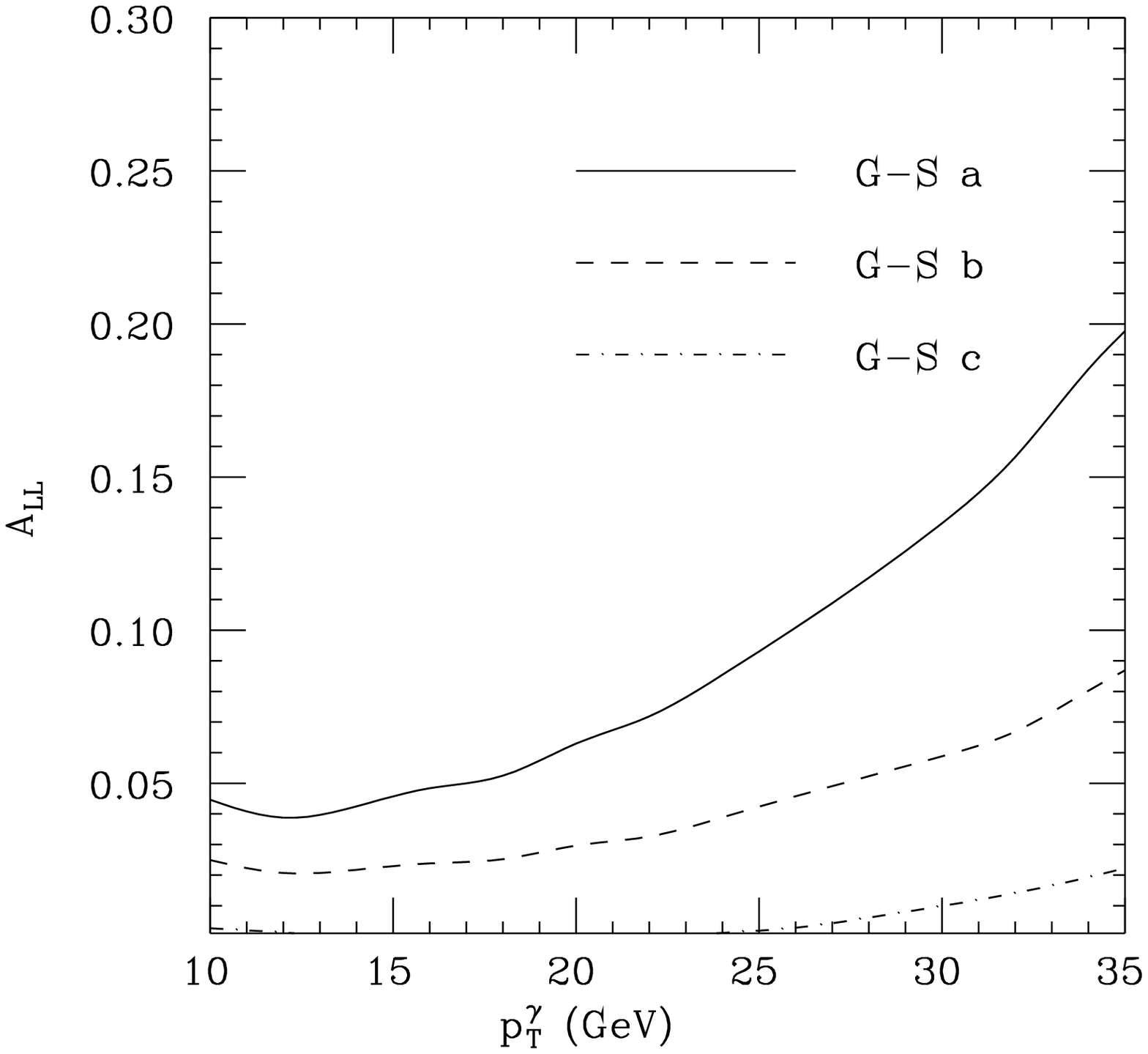}}
\centerline{a)\hspace{6cm} b)}
\caption{}
\end{figure}
\begin{figure}
\epsfxsize=60mm
\centerline{\epsfbox{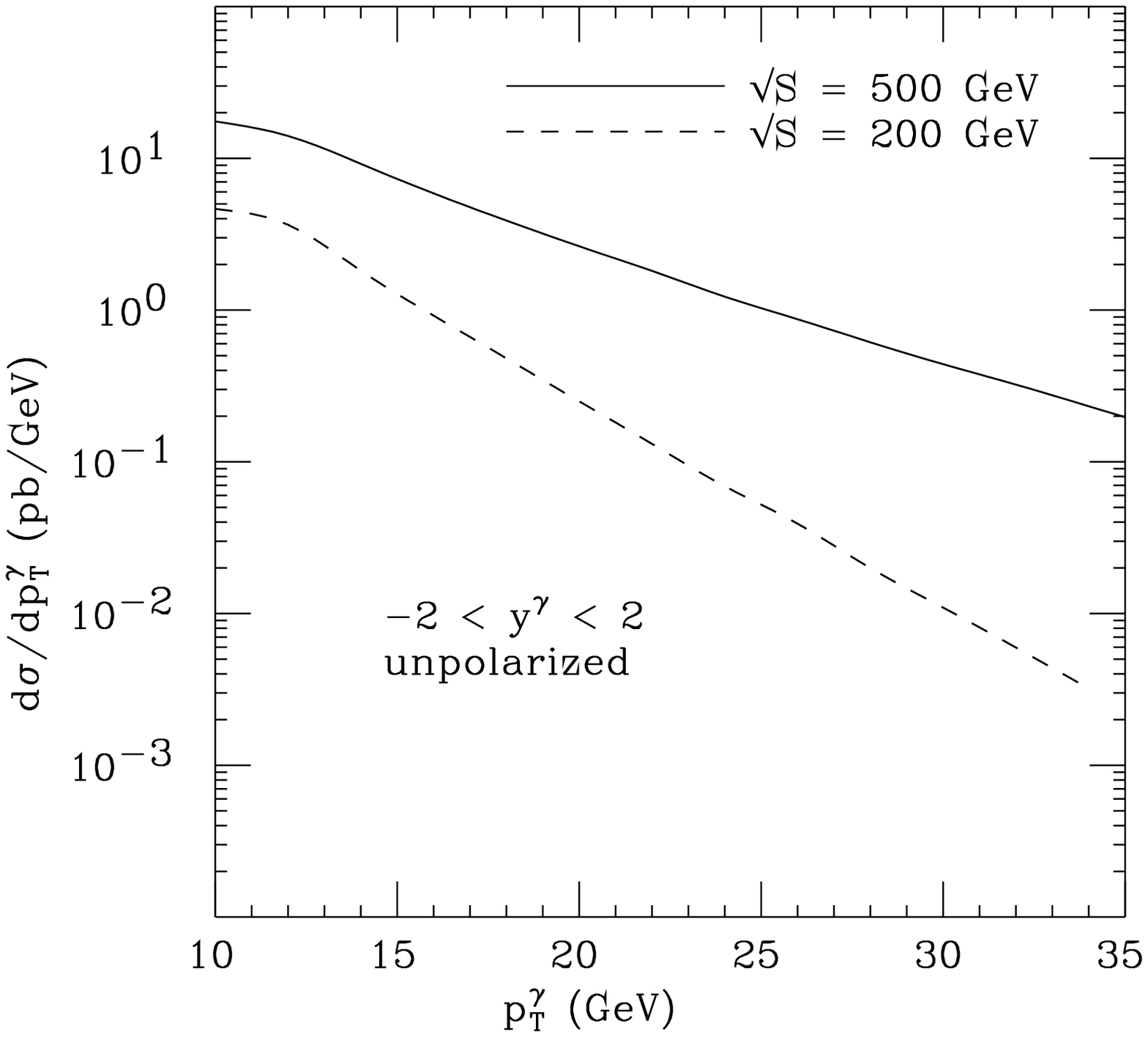}
\epsfxsize=60mm
\epsfbox{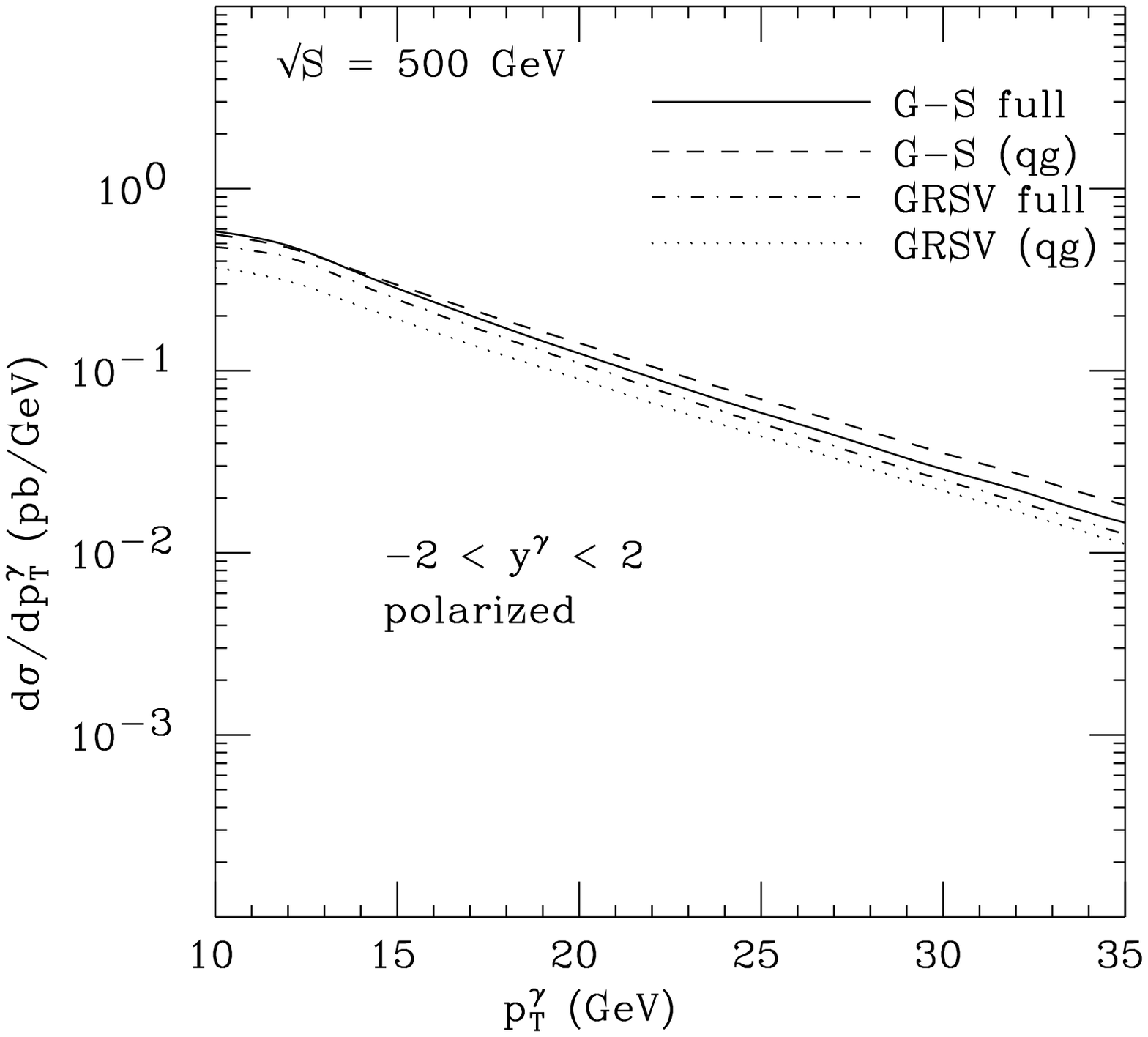}}
\centerline{a)\hspace{6cm} b)}
\caption{}
\end{figure}
In Figs.1a and 1b we display the unpolarized 
and the polarized cross sections. The results indicate that it
will be difficult to measure this cross section beyond about
$p_T=20$ GeV, even with the planned high luminosities at RHIC. More
interestingly, Fig.1b suggests,as we expected, that GRSV distributions give 
very similar
results and cannot be separated by this process.

In Figs.2a and 2b we show the longitudinal asymmetries as predicted by the
various parameterizations. The asymmetry, $A_{LL}$, is defined by the
relation
\begin{equation}
A_{LL}=\frac{ \frac{d\Delta\sigma}{dp_T} }{\frac{d\sigma}{dp_T} },
\end{equation}
the ratio of the polarized to the unpolarized cross section, and gives a
measure of the spin dependence or spin sensitivity of the process. 
Again the two GRSV distributions, as expected, give
similar results, while the G-S ones give clearly distinguishable
results. The predictions also indicate that the asymmetries are not very
large in the measurable region of the cross section, varying between $5$
and $8\%$ for the GRSV and $4$ and $6\%$ for the G-S $a$
parameterization. 

Fig.2b suggests, on the
other hand, that the three G-S distributions give substantially different
predictions, and will probably be distinguishable. The G-S $a$
parameterization gives similar results to the two GRSV versions.

It will require very high statistics measurements to
measure this asymmetry, but this may be achievable at RHIC, as long as
the cross section can be measured, given the planned detectors. 

At $\sqrt{S}=500$ GeV the situation improves somewhat with regards to the
size of the cross section. Fig.3a compares the $p_T$ distributions at
$\sqrt{S}=500$ and $200$ GeV for the unpolarized cross section. As
expected, at higher cms energies the cross section is substantially
larger. In Fig.3b we show the polarized cross section at
$\sqrt{S}=500$ GeV. On the figure we also show the full prediction given by
the G-S $a$ distribution as well as the contribution form $q g$
initiated subprocesses. In the case of G-S $a$ the contribution for the
$q\bar{q}$ initiated subprocess (not shown) is negative. Corresponding
distributions are shown for the GRSV 1 parameterization. In this case the
$q\bar{q}$ process gives a positive contribution but it is substantially
smaller than the $q g$ one. The G-S $a$ and GRSV 1 distributions predict
similar cross sections, but the relative importance of the subprocesses
is clearly very different. The most interesting aspect of this result
from the point of view of sensitivity to $\Delta G$ is the fact that in
both cases, the $q g$ initiated process dominates. We should mention
that in this calculation we do not include contributions for the higher
order process ($O(\alpha_s^2)$) $gg\rightarrow \gamma\gamma$, preferring to keep consistently to $O(\alpha_s)$.

\section{Conclusions}
We have presented a complete parton level study of the NLO radiative 
corrections to Drell Yan for a non zero $q_T$ of the lepton pair, thereby 
extending the result of ref.~\cite{EMP} which discussed the unpolarized case. 
Our results completely overlap with these former calculations, 
except for the extraction of the helicity violating contributions. The derivation of these terms requires a complete recalculation of all the diagrams in a general way. Numerical results in the case of double photon have been given. 
The cross section is one power of $\alpha_{em}$ suppressed compared to the single photon case. Around $\sqrt{s}\sim 400-500$ GeV the corrections 
are sizeable with a $K-$factor $K_f\sim 2$.

\end{document}